\documentclass[12pt]{iopart}

\usepackage{iopams}
\expandafter\let\csname equation*\endcsname\relax
\expandafter\let\csname endequation*\endcsname\relax
\usepackage{amsmath}
\usepackage{hyperref}
\usepackage{manuscriptStyleFileTia}
\usepackage[version=4]{mhchem}
\usepackage{orcidlink}
\usepackage{biblatex}

\usepackage{lineno}

\date{\today}

\begin{document}

\title[Remote Sensing]{A physics-aware data-driven surrogate approach for fast atmospheric radiative transfer inversion}

\author{Cristina Sgattoni$^{1,4}$\orcidlink{0000-0001-5734-0856}
, Luca Sgheri$^2$\orcidlink{0000-0002-6014-9363}, Matthias Chung$^3$\orcidlink{0000-0001-7822-4539}}

\address{$^1$Istituto per la BioEconomia (IBE), Consiglio Nazionale delle Ricerche (CNR), Via Madonna del Piano, 10, Sesto Fiorentino, 50019, Firenze, Italy}

\address{$^2$Istituto per le Applicazioni del Calcolo 
(IAC), Consiglio Nazionale delle Ricerche (CNR), Via Madonna del Piano, 10, Sesto Fiorentino, 50019, Firenze, Italy}

\address{$^3$Department of Mathematics, Emory University, 400 Dowman Drive, Atlanta, GA, USA}

\address{$^4$INdAM Research Group GNCS, P.le Aldo Moro, 5, 00185 Roma, Italy}

\eads{\mailto{cristina.sgattoni@cnr.it}, \mailto{luca.sgheri@cnr.it}, \mailto{matthias.chung@emory.edu}}

\vspace{10pt}
\begin{indented}
\item[]\today
\end{indented}

\begin{abstract}
FORUM (Far-infrared Outgoing Radiation Understanding and Monitoring) was selected in 2019 as the ninth Earth Explorer mission by the European Space Agency (ESA). Its primary objective is to collect interferometric measurements in the Far-InfraRed (FIR) spectral range, which accounts for 50\% of Earth's outgoing longwave radiation emitted into space, and will be observed from space for the first time. Accurate measurements of the FIR at the top of the atmosphere are crucial for improving climate models. Current instruments are insufficient, necessitating the development of advanced computational techniques. FORUM will provide unprecedented insights into key atmospheric parameters, such as surface emissivity, water vapor, and ice cloud properties, through the use of a Fourier Transform Spectrometer. To ensure the quality of the mission’s data, an End-to-End Simulator (E2ES) was developed to simulate the measurement process and evaluate the effects of instrument characteristics and environmental factors.

The core challenge of the mission is solving the retrieval problem, which involves estimating atmospheric properties from the radiance spectra observed by the satellite. This problem is ill-posed and regularization techniques are necessary to stabilize the solution.
In this work, we present a novel data-driven approach to approximate the inverse mapping for the retrieval problem, aiming to achieve a solution that is both computationally efficient and accurate.

In the first phase, we generate an initial approximation of the inverse mapping using only simulated FORUM data. In the second phase, we improve this approximation by introducing climatological data as a priori information and using a neural network to estimate the optimal regularization parameters during the retrieval process.

While our approach does not match the precision of full-physics retrieval methods, its key advantage is the ability to deliver results almost instantaneously, making it highly suitable for real-time applications. Furthermore, the proposed method can provide more accurate a priori estimates for full-physics methods, thereby improving the overall accuracy of the retrieved atmospheric profiles.

\end{abstract}

\vspace{2pc}
\noindent{\it Keywords}: radiative transfer inversion, data-driven inversion, regularization, parameters estimation, physics-guided machine learning

\submitto{Inverse Problems}

\section{Introduction}\label{sec:introduction}
The composition of Earth's atmosphere significantly influences how heat is transferred, making it crucial for accurate climate and weather predictions \cite{simmons2010,ipcc2021,noaa2022}.
The \emph{radiative transfer equation} models the intensity of electromagnetic radiation as it passes through a medium. For homogeneous media, solutions of this equation can be found analytically by combining Lambert-Beer and Planck's law. However, for heterogeneous media, like Earth's atmosphere, solutions require numerical integration techniques~\cite{chandrasekhar1960}. 

Interferometers, measuring radiation intensity as a function of wavelength (the spectrum), are operationally located on satellites. Along the instrument's line of sight, a high-resolution spectrum as a result of the radiative transfer is recorded. The objective is to use these spectra to reconstruct the composition of atmospheric gases and, subsequently, infer Earth's heat balance.
\\

In September 2019, the European Space Agency (ESA) announced the \emph{Far-infrared Outgoing Radiation Understanding and Monitoring} (FORUM) mission as its ninth Earth Explorer mission. Scheduled for launch in 2027, FORUM will be equipped with a Fourier Transform Spectrometer (FTS) interferometer, which will point towards the nadir, i.e., along the vertical from the satellite to the Earth's surface, to gather spectral data for analyzing Earth's heat signature, see left panel of Figure~\ref{fig:figure2}. The FORUM instrument will work from 100 to 1600~\si{cm^{-1}}, furnishing spectrally resolved radiances from space in the far-infrared band for the first time. Once in orbit, FORUM will generate a significant amount of data that will need to be processed and analyzed. For instance, the state-of-the-art Infrared Atmospheric Sounding Interferometer -- New Generation (IASI-NG) spectrometer \cite{Eumetsat} will output $16$ spectra every $800$ \rm{ms}, with some short pauses for calibration, resulting in about 1,382,400 spectra per day. It is expected that just the FORUM spectrometer alone will generate more than 10,000 spectra per day. Before the FORUM mission launch, an End-to-End Simulator (E2ES) for the FORUM spectrometer was developed during the early stages of mission development. This critical tool provides a proof-of-concept for the instruments and allows for a rigorous evaluation of how instrument characteristics and scene conditions will impact the quality of retrieved measurements, see \cite{palchetti2020,sgheri2022} for details.

The atmospheric spectrum depends on a set of factors that can be collectively referred to as the \emph{atmospheric state} and includes surface temperature, surface emissivity, the vertical profiles of temperature, and concentrations of the atmospheric gases. After discretization, the atmospheric state can be denoted by $\bfx\in\mathbb{R}^n$, where $n$ is the number of parameters considered. Let $\bfF:\mathbb{R}^n \to \mathbb{R}^q$ represent the combination of the radiative transfer and instrumental effects, $\bfepsilon \in \mathbb{R}^q$ be some additive noise in the data acquisition process and $\bfy \in \mathbb{R}^q$ be the spectrum, with $q$ number of spectral points, then the relationship between atmospheric state, radiative transfer, and observations can be expressed as
$$
\bfF(\bfx) + \bfepsilon = \bfy.
$$
Two equally important tasks are associated with radiative transfer: the direct problem -- determining $\bfy$ from $\bfx$ -- and the inverse problem, determining $\bfx$ from $\bfy$. Both are of interest for various applications. The direct problem (also referred to as the forward problem) is used both as a sub-problem of the inverse problem and for the assimilation of the radiances in climatological and meteorological models, see for instance \cite{noh2017,dellafera2023} and the bibliography therein. The inverse problem is also referred to as the \emph{retrieval problem}, and it is used to monitor the atmosphere. On this subject, the bibliography is extensive. Good introductory monographs on the subject are \cite{liou2002,wallace2006,zdunkowski2007,stamnes2017}, while the classical book \cite{rodgers2000} focuses on error analysis and the solution of the inverse problem via regularization techniques. On the regularization, see also \cite{doicu2010}. The direct problem, as well as the inverse problem (in which the direct problem is a sub-problem), can be solved using a full-physics method. However, the computational cost of this approach is too high to achieve Near Real-Time (NRT) data analysis.

To solve this problem, different approaches have been proposed to speed up the computation of radiative transfer. The Radiative Transfer for TOVS (RTTOV)~\cite{saunders2018} (along with PC-RTTOV~\cite{matricardi2010}, the principal component version of RTTOV) and sigma-IASI~\cite{amato2002} are two candidates for fast radiative transfer codes. The sigma-IASI code, which is based on the parameterization of the influence of temperature and the concentrations of atmospheric gases on radiative transfer, has recently been extended to include the FORUM and IASI-NG frequency ranges, as well as cloud parametrization~\cite{martinazzo2021}.

In this work, we develop a novel generalized data-driven approach to approximate the inverse mapping ``$\bfF^{-1}$'' to obtain a computationally fast approximation of the solution to the retrieval problem. We utilize the end-to-end simulated measurements from the FORUM spectrometer to generate training data. The key point of our work is the inclusion of climatological data. Since the inverse problem is ill-posed (see for instance, \cite{rodgers2000}), any approach that does not include regularization techniques is not able to retrieve accurate profiles. The usual approach to solve the inverse problem is to use climatological or correlative data as a priori information. Climatological data, however, depends on the particular case under analysis, so it is not possible to include this information in the training stage of a machine-learning approach. Our solution is to use an additional neural network to find the optimal Tikhonov parameters to be used in the regularization procedure. While this approach does not guarantee a solution of the same accuracy as full-physics retrieval methods, the solution is produced almost instantaneously. If applied to climatology, the provided solution can yield a priori estimates that are statistically closer to the true atmospheric state. Using a priori values that are closer to the true ones generally provides an advantage in the determination of the solution of the inverse problem \cite{merchant2020,sgheri2024}, because it reduces the bias in the solution introduced by errors in the a priori. 

This work is organized as follows. In \Cref{sec:background}, we provide an overview of the physics behind the forward problem (\Cref{sub:direct}) and the classical inversion framework (\Cref{sub:inverse}). In \Cref{sec:solsch}, we introduce our surrogate inversion scheme, beginning with the purely data-driven phase (\Cref{sub:datadriven}), followed by the inclusion of a priori information (\Cref{sub:aprioriapp}), and the estimation of regularization parameters (\Cref{sub:lambdaopt}). In \Cref{sub:offscheme}, we summarize and illustrate the entire solution scheme, encompassing both the offline and online components. In \Cref{sec:dataorg}, we describe the structure of the data. In \Cref{sec:results}, we present and analyze the results of our numerical experiments, covering the data-driven phase (\Cref{sub:datadrivenres}), the impact of incorporating a priori information (\Cref{sub:aprioriappres}), and a comparison with a full-physics approach (\Cref{sub:fullph}). Finally, in \Cref{sec:concl}, we summarize our contributions, discuss further implications, and suggest directions for future work.

\section{Background}\label{sec:background}

\subsection{Radiative Transfer}\label{sub:direct}
In the following, we briefly review the underlying physics of the atmospheric remote sensing problem. For an atmospheric \emph{homogeneous} layer, given a wavenumber $\nu$, the radiative transfer equation can be written as
\begin{equation}\label{eq:rt1}
    \frac{\d I_\nu}{\d z}(z) = -\alpha_\nu(p,T,c) I_\nu(z) + \alpha_\nu(p,T,c) B_\nu(T), \qquad \mbox{with } \quad I_\nu(z_0)=I_{\nu_0},
\end{equation}
where $z$ denotes the altitude, $I_\nu(z)$ is the intensity of radiation, $B$ represents the Planck function, and $\alpha_\nu(p,T,c)$ corresponds to the attenuation coefficient, depending on the layer pressure $p$, the layer temperature $T$, and the layer concentration of atmospheric gases $c$.
The attenuation coefficient can be obtained using a spectroscopic database, such as the HIgh resolution TRANsmission (HITRAN) 2020 database, see \cite{gordon2022}.
Equation~\eqref{eq:rt1} is derived by combining both Lambert-Beer's and Planck's laws. The analytic solution of the initial value problem~\eqref{eq:rt1} is given by
\begin{equation}\label{eq:rt_sol}
    I_\nu(z) = I_{\nu_0}\e^{-\alpha_\nu(p,T,c)(z-z_0)} + B_\nu(T)\left(1-\e^{-\alpha_\nu(p,T,c)(z-z_0)}\right).
\end{equation}

\begin{figure}
\begin{center}
\includegraphics[width=0.30\textwidth]{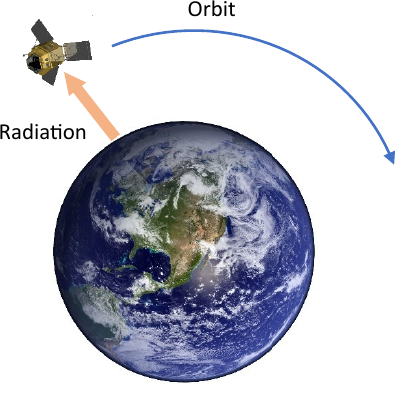}
\includegraphics[width=0.48\textwidth]{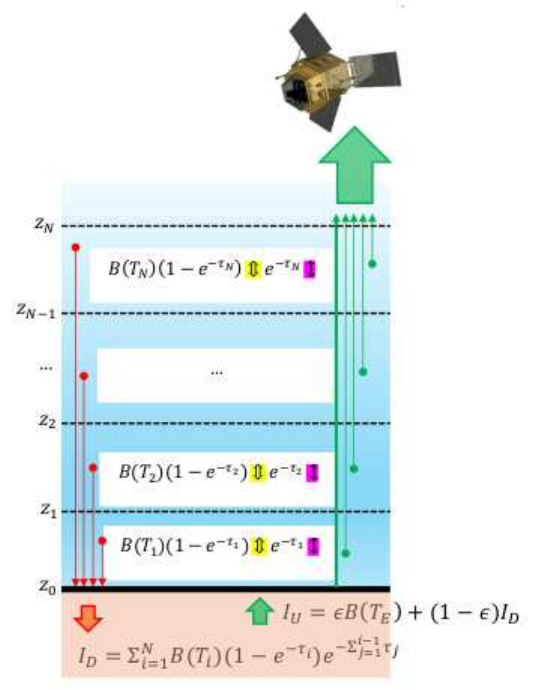}
\caption{Left panel: FORUM orbit with upwelling radiation seen by the instrument. Right panel: Radiative transfer and the discretization of the atmosphere. The downwelling radiation is indicated in red and the upwelling radiation is in green. In each atmospheric layer, the attenuation factor is highlighted with the purple arrow, while the thermal emission is highlighted with the yellow arrow.}
\label{fig:figure2}
\end{center}
\end{figure}

Since the atmosphere is not a homogeneous medium, it is discretized in layers that are considered homogeneous, see right panel of \Cref{fig:figure2}. Let $z_0,\ldots,z_N$ the vertical discretization of the atmosphere, with $z_0$ being Earth level and $z_N$ the altitude of the observer, or the atmosphere limit if the instrument is outside the atmosphere. In each layer $[z_{i-1},z_{i}]$, $T_i$ is the Curtis-Godson \cite{godson1953} average temperature of the layer, $\alpha_i$ and $\tau_i=\alpha_i(z_i-z_{i-1})$ are the frequency-dependent attenuation coefficient and optical depth, respectively. For clarity, we drop the explicit frequency dependence in the notation from now on.

According to (\ref{eq:rt_sol}), the electromagnetic radiation passing through each  layer $i$ is attenuated by a factor $\e^{-\tau_i}$. This energy is absorbed and re-emitted as $(1-\e^{-\tau_i})B(T_i)$ isotropically according to Planck law. The same mechanism is also valid for Earth's surface. The coefficient $\epsilon$ that measures the fraction of radiation absorbed is called spectral emissivity and depends on the frequency and the type of soil on the surface. Earth's surface absorbs a fraction $\epsilon$ of the radiation $I_D$ reaching it and re-emits it according to Planck law $\epsilon B(T_E)$, where $T_E$ is Earth's surface temperature. The remaining fraction $(1-\epsilon)I_D$ is reflected back into space, so that source upwelling radiation becomes: $I_U=\epsilon B(T_E)+(1-\epsilon)I_D$. To calculate the downwelling radiation $I_D$ we must factor the contributions of the emission of all the layers, each attenuated by passing through the lower layers, resulting in $\sum_{i=1}^{N} B(T_{i}) (1 - \e^{-\tau_{i}}) \e^{-\sum_{j=1}^{i-1} \tau_{j}}$.

Analogously, to calculate the upwelling radiation reaching the instrument we must factor the attenuation of the source radiation $I_U$ by all the atmospheric layers. To this expression, we must add the emission of each layer attenuated by the passage from the upper layers. The final discretized form of the radiative transfer equation is \cite{zdunkowski2007}:
\begin{gather}\label{eq:RTE_discr}
\begin{split}
I(z_{N})=\left( \epsilon \ B(T_{E}) + (1-\epsilon) \left( \sum_{i=1}^{N} B(T_{i}) \ (1 - \e^{-\tau_{i}}) \ \e^{-\sum_{j=1}^{i-1} \tau_{j}} \right) \right) \e^{-\sum_{i=1}^{N} \tau_{i}}\\
+\sum_{i=1}^{N} B(T_{i}) \ (1 - \e^{-\tau_{i}}) \ \e^{-\sum_{j=i+1}^{N} \tau_{j}}. 
\end{split}
\end{gather}

The atmospheric unknowns of (\ref{eq:RTE_discr}) appear in the attenuation coefficients making the retrieval of atmospheric gases a challenging, ill-conditioned problem.

\subsection{Inversion problem}\label{sub:inverse}
The retrieval problem can be stated as a Bayesian inference problem where the aim is to obtain a posterior distribution of the atmospheric parameters $\bfx$, given observations $\bfy$ and prior information of the atmospheric parameters; the posterior is given by
$$
\pi(\bfx \mid \bfy) = \frac{\pi(\bfy \mid \bfx) \ \pi(\bfx) }{\pi(\bfy)}\propto  \pi(\bfy \mid \bfx) \ \pi(\bfx).
$$
In this remote sensing problem, it is commonly assumed that the measurement noise and the prior follow a Gaussian distribution \cite{rodgers2000}. More precisely, we assume $\bfepsilon = \bfy -\bfF(\bfx) \sim \calN({\bf0},\bfS_\bfy)$ and $\bfx \sim \calN(\bfx_a,\bfS_a)$ with symmetric positive definite covariance matrices $\bfS_\bfy$ and $\bfS_a$. The radiative transfer discussed in \Cref{sub:direct} is discretized and mapped onto the data by the forward process $\bfF$. The matrix $\bfS_\bfy$ can be explicitly estimated from the instrumental characteristics, see for instance \cite{crevoisier2014} for the IASI-NG sounder. The mean $\bfx_a$ can be determined for instance from correlated measurements or climatology. In both cases, it is also possible to obtain an estimate of $\bfS_a$, that represents either the correlated measurements error or the atmospheric variability. In this remote sensing application, the maximum a posteriori (MAP) estimate is given by $\bfx_{\rm map} \in \argmax_{\bfx} \pi(\bfx \mid \bfy)$ and can be computed by solving the optimization problem
\begin{gather}
\bfx_{\rm map} = \argmin_\bfx\chi^{2}(\bfx), \qquad \mbox{where}\label{eq:oe}\\
\chi^{2}(\bfx)=\thf (\bfy-\bfF(\bfx))\t\bfS_{y}^{-1}(\bfy-\bfF(\bfx)) \, + \, \thf(\bfx_{a}-\bfx)\t\bfS_{a}^{-1}(\bfx_{a}-\bfx).\label{eq:chi2}
\end{gather}
Details on its derivations can be found for instance in \cite{bardsley2018computational,calvetti2007introduction}. This radiative transfer inversion problem using full-physics methods \Cref{eq:oe} may also be interpreted as a nonlinear Tikhonov regularized variational inverse problem and is referred to as the Optimal Estimation (OE) method in the field of atmospheric retrieval problems and introduced by Rodgers \cite{rodgers2000,engl1996regularization}. Here, other related techniques include methods based on principal component analysis \cite{amato2002,matricardi2010}. By defining $\bfL_\bfy$ and $\bfL_a$ as symmetric decompositions of the precision matrices $\bfS_\bfy^{-1} = \bfL_\bfy\t\bfL_y$ and $\bfS_a^{-1} = \bfL_a\t\bfL_a$, e.g., by Cholesky or eigenvalue factorization, definition~\eqref{eq:chi2} can be restated as :
\begin{equation}\label{eq:chi2_a}
\chi^{2}(\bfx)=\ \thf \norm[2]{\bfL_\bfy(\bfF(\bfx) -\bfy)}^2 + \thf \norm[2]{\bfL_a(\bfx- \bfx_a)}^2.
\end{equation}
Solving \Cref{eq:oe} using full-physics methods is computationally demanding because the attenuation coefficients $\alpha_\nu$ from \Cref{eq:RTE_discr} must be evaluated for each gas, each layer, and each wavelength using the spectroscopic database. As a result, the evaluation of $\bfF$ is time-consuming \cite{clough05,klima2014}. This process must be repeated at each iteration because the $\alpha_i$ depends both on temperature and the concentrations of the atmospheric gases, that are normally targets of the retrieval. To solve  \Cref{eq:oe}, Gauss-Newton-type methods are often employed because they normally need a limited number of forward model evaluations. However, due to the non-convexity and heterogeneous sensitivities of $\chi^{2}$ convergence problems may arise. To address this, Levenberg-Marquardt techniques, which have regularizing properties, are frequently applied  \cite{dinelli2021}. Still, full-physics methods are not adequate to cope with the amount of data produced by modern instruments, making the development of alternative fast and accurate inversion techniques essential.

\section{Surrogate inversion scheme} \label{sec:solsch}

Due to the computational challenges and bottlenecks laid out in the previous section, we propose a novel physics-aware data-driven approach to overcome these challenges in the atmospheric retrieval problem. In the first step, we directly approximate the inverse model ``$\bfF^{-1}$'' using a linear surrogate model, while in the second step, we propose to utilize deep learning methods to improve our predictions by incorporating additional physical information. In \Cref{sub:datadriven}, we introduce a fully data-driven method that is subsequently incorporated with a priori information in \Cref{sub:aprioriapp}. Following this, in Section \Cref{sub:lambdaopt} we propose a procedure for estimating the regularization parameters using a neural network. In the final Section \Cref{sub:offscheme} we summarize the entire solution chain.

\subsection{Data-driven surrogate model}\label{sub:datadriven}

Reduced order models have been well-established for dynamical systems and there exists a vast literature on approaches to build reliable surrogate forward models. We point the interested reader to the book \cite{antoulas2020interpolatory} and references within. Another obvious approach is to use a linear approximation, e.g., using a Taylor expansion $\bfF(\bfx) \approx \bfF(\bfx_0) + ({\rm D}_\bfx \bfF(\bfx_0))\bfx$ for a given $\bfx_0$ to simplify the model simulation. Here, we follow a different approach. Since we are interested in the inverse problem, we aim \emph{not} to approximate $\bfF$ but to approximate the action of its inverse mapping $\bfy \mapsto \bfx$. We aim to find a linear operator $\bfZ$ that minimizes the \emph{Bayes risk}, i.e.,
\begin{equation}\label{eq:empRisk}
    \min_{\bfZ} \quad \bbE_{\bfx,\bfy} \ \norm[\bfS_\bfy^{-1}]{\bfx -\bfZ\bfy}^2,
\end{equation}
where $\bbE_{\bfx,\bfy}$ denotes the expectation over $\bfx$ and $\bfy$ and for symmetric positive matrices $\bfGamma$ we define $\norm[\bfGamma]{\bfx} = (\bfx\t \bfGamma \bfx)^{1/2}$. Even if the distribution of $\bfx$ is known, obtaining the distribution $\bfy = \bfF(\bfx) + \bfepsilon$ is computationally intractable. Assume we are given data pairs $\{\bfx_j,\bfy_j\}_{j=1}^m$, with $m$ number of observations. We obtain through a simulated forward propagation $\bfy_j = \bfF(\bfx_j) +\bfepsilon_j$, then with the notation $\bfX = [\bfx_1,\ldots,\bfx_m]$, $\bfY = [\bfy_1,\ldots,\bfy_m]$, the \emph{empirical Bayes risk} is given by
\begin{equation}\label{eq:empRisk2}
  f(\bfZ) =  \tfrac{1}{m}\norm[\fro]{\bfL_\bfy(\bfX -\bfZ\bfY)}^2,
\end{equation}
where $\norm[\fro]{\mdot}$ denotes the Frobenius norm. A minimizer of~\eqref{eq:empRisk} can found by determining stationary points of $f$, i.e., $\frac{\partial f(\bfZ)}{\partial \bfZ} = {\bf0}$. Noticing that 
\begin{equation}
   m f(\bfZ) = \trace{\bfX\t\bfL_\bfy\t\bfL_\bfy\bfX}-2\trace{\bfX\t\bfL_\bfy\t\bfL_\bfy\bfZ\bfY} + \trace{\bfY\t\bfZ\t\bfL_\bfy\t\bfL_\bfy\bfZ\bfY}
\end{equation}
where $\trace{\mdot}$ denotes the trace of a matrix, we may compute its derivative as
\begin{align}
m\frac{\partial f}{\partial \bfZ} &= -2\bfL_\bfy\t\bfL_\bfy\bfX\bfY\t + 2 \bfL_\bfy\t\bfL_\bfy\bfZ\bfY\bfY\t
\end{align}
and a minimizer $\widehat\bfZ$ of $f$ solves $\widehat\bfZ\bfY\bfY\t= \bfX\bfY\t$.  
The (minimal norm) solution to this equation is given by 
\begin{equation}\label{eq:fullmi}
    \widehat\bfZ = \bfX\bfY^\dagger,
\end{equation}
where $\dagger$ is the Moore-Penrose pseudoinverse. 

Let us provide a couple of remarks. Certainly, the empirical Bayes risk approach can be used to obtain an approximation on the forward operator $\widehat\bfA\bfx \approx \bfF(\bfx)$, however, due to the ill-posed nature of the inverse process we observe that using $\widehat \bfA^\dagger$ is inferior to directly approximating the inverse operation $\widehat \bfZ\bfy$. In large-scale and severely ill-posed settings further rank constraints on $\bfZ$ may be considered as developed in \cite{chung2017optimal}, however, in the remote sensing application at hand $\bfZ$ is of moderate dimension and rank constraints are unnecessary. Here the fast repeated application of $\widehat\bfZ$ is of more concern.

Assume we have computed $\widehat \bfZ$ given training data $\left\{\bfx_j,\bfy_j\right\}_{j = 1}^m$ in an off-line phase. If new data $\bfy$ becomes available we just need a fast matrix-vector multiplication to obtain a prediction of the atmospheric state $\widehat\bfx$, i.e.,
\begin{equation}\label{eq:ddsol}
    \widehat\bfx = \widehat\bfZ\bfy.
\end{equation}
Notice this purely data-driven approach does not include any prior knowledge of the atmospheric state, hence our next step is to improve upon the atmospheric state estimate to include the prior in the form of regularization. 

\subsection{Data-driven approach with a priori correction}\label{sub:aprioriapp}

Data-driven methods like neural networks have revolutionized science and our daily lives. While powerful for reconstruction and prediction, data-driven methods can also struggle to predict broader system dynamics or generate physically unrealistic parameters.

New approaches integrate underlying physics and prior knowledge into data-driven prediction processes. These approaches have demonstrated improving reliability and predictions \cite{raissi2017physics, genedy2024physics}. Focusing on the inverse problem at hand, we incorporate the prior information to improve upon the data-driven prediction. We assume that superior reconstruction can be obtained by balancing between the data-driven prediction $\widehat\bfx$ and the prior $\bfx_a$, i.e., 
\begin{equation}\label{eq:ap}
    \bfx_{\bflambda} = \argmin_{\boldsymbol{\xi}}  \quad \thf \norm[2]{\bfL_\bfx(\boldsymbol{\xi} - \widehat\bfZ\bfy) }^2 + \thf \norm[2]{\bfLambda \bfL_a (\boldsymbol{\xi} - \bfx_a ) }^2,
\end{equation}
where $\bfLambda = \diag{\bflambda}$ is a positive diagonal weighting matrix representing the importance we decide to assign to the a priori contribution, with $\bflambda$ indicating its diagonal, and $\bfL_\bfx$ is the symmetric decomposition of the precision matrix inverse $\bfS_\bfx^{-1} = \bfL_\bfx\t\bfL_\bfx$. Normally, the precision matrix of the unknowns, $\bfS_\bfx$, is estimated through the linear propagation of errors \cite{rodgers2000}. However, since we do have the solutions $\bfx_j$ for the $m$ points of the training set, we can use the approximation
\begin{equation}(\bfS_\bfx)_{kl}\approx\frac{1}{m}\sum_{j=1}^m(\widehat x^{k}_j- x_j^k)(\widehat x^{l}_j- x_j^l).
\end{equation}
The unique solution of~\eqref{eq:ap} is given by
\begin{equation}\label{eq:apsol}
    \bfx_\bflambda = \left(\bfL_\bfx + \bfL_a\t\bfLambda^2\bfL_a \right)^{-1} \left( \bfL_\bfx\widehat\bfZ\bfy+\bfL_a\t\bfLambda^2\bfL_a\bfx_a\right).
\end{equation}
Note, when $\bflambda = \mathbf{0}$, the solution in~\eqref{eq:apsol} coincides with the purely data-driven solution in~\eqref{eq:ddsol}.
Typically, a regularization technique includes a term of the form $\norm[2]{\bfL_\bfy(\widehat\bfA\bfx -\bfy)}^2$, which is computed in the measurement space and may be affected by the ill-posedness of the inverse problem. In contrast, the minimization of \Cref{eq:ap} is performed in the parameter space, which in this application has a much lower dimensionality. This minimization balances the solution’s distance from the data-driven result with its distance from the prior. The vector $\bflambda$  determines the weighting between these two terms and the selection of a regularization $\bflambda$ can be cumbersome. Here, we select $\bflambda$ by utilizing a neural network described in \Cref{sub:lambdaopt} and discussed in \Cref{sec:results}. 

\subsection{Regularization parameters estimation}\label{sub:lambdaopt}

There are well-established statistical techniques to determine the regularization vector $\bflambda$, such as the \emph{unbiased predictive risk estimator} (UPRE), the \emph{discrepancy principle} (DP), \emph{generalized cross validation} (GCV), see \cite{bardsley2018computational}. While these methods perform well in practice,  they all share a common drawback: the associated computational cost often exceeds that of solving the inverse problem itself. This issue is exacerbated when multiple regularization parameters are involved, as in our case. To address this, we propose a data-driven approach for estimating regularization parameters. Specifically, following the method developed in \cite{afkham2021learning}, we train a deep neural network to predict these parameters using a training set of $m_{t_1}$ cases.
An important part of the training data consists of the optimal regularization parameters $\{\bflambda_j^{\rm opt}\}_{j=1}^{m_{t_1}}$ for the $m_{t_1}$ training points. To obtain these, we employ a deterministic approach.
In particular, we formulate a bilevel optimization problem. The outer problem is as follows,
\begin{equation}\label{eq:tf}
    \bflambda_j^{\rm opt} = \argmin_{\bflambda \geq {\bf0}} \ \norm[2]{(\bfx_{\bflambda})_j - \bfx_j}
\end{equation}
where $(\bfx_{\bflambda})_j$ is the solution of the inner problem described in Equation~\eqref{eq:apsol} for case $j$ in the training set.

The remaining training data consist of the prior estimates $\{(\bfx_a)_j\}_{j=1}^{m_{t_1}}$ and the data-driven solutions $\{\widehat\bfx_j\}_{j=1}^{m_{t_1}}$. 
We assume there exists a well-defined mapping from generic vectors $\widehat\bfx$ and $\bfx_a$ to $\bflambda$. Specifically, we define the mapping as $\widetilde\bfPhi: \bbR^n\to \bbR^n$, where $\widetilde\bfPhi(\widehat\bfx - \bfx_a) = \bflambda$. 
To reduce the complexity of the mapping, we condense the first two pieces of information by using the difference between $\widehat\bfx$ and $\bfx_a$. 
We then approximate this mapping with a neural network $\bfPhi:\bbR^n \times \bbR^k \to \bbR^r$ parameterized by $\bftheta$,  i.e., 
\begin{equation}\label{eq:ffnn1}
    \bfPhi\big((\widehat \bfx - \bfx_a), \bftheta\big) = \bflambda.
\end{equation}

Given training data $\Big\{\big(\widehat\bfx_j - (\bfx_a)_j\big), \bflambda_j^{\rm opt}\Big\}_{j = 1}^{m_{t_1}}$, we determine the network parameters $\bftheta$ by solving
\begin{equation}\label{eq:ffnn}
\bftheta = \argmin_{\boldsymbol{\zeta}} \ \tfrac{1}{m_{t_1}} \sum_{j = 1}^{m_{t_1}} \norm{\bfPhi\Big(\big(\widehat \bfx_j - (\bfx_a)_j\big),\boldsymbol{\zeta}\Big)- \bflambda_j^{\rm opt}}.
\end{equation}

A key novelty of our approach compared to \cite{afkham2021learning} is that we do not learn a mapping from the data $\bfy_j$ to the regularization parameters $\bflambda_j$ but rather from the estimated parameter $\widehat \bfx_j$ to the regularization parameter so that again we operate within the parameters domain. The advantage of this approach is the reduced parameter size of the neural network which is easier to solve.  

To improve the numerical stability and allow the network to predict a wider range of values more effectively, the neural network output is not the $\bflambda$ vector itself, but its logarithm. 

\subsection{Solution scheme}\label{sub:offscheme}
\begin{figure}[H]
\begin{center}
\includegraphics[width=1.
\textwidth]{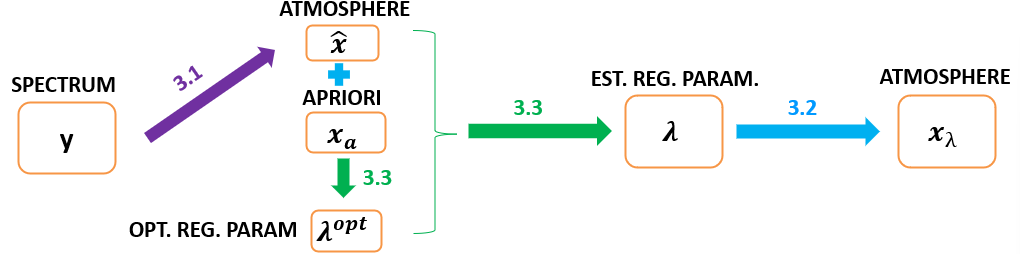}
\caption{Solution scheme where the different steps are represented with arrows of different colors with the related section indication: purple for data-driven step, blue for regularization step and green for regularization parameters computation and estimation.}\label{fig:scheme}
\end{center}
\end{figure}

Here we summarize the complete solution scheme, as described in the previous sections and depicted in Figure \ref{fig:scheme}.

The solution process begins with a fully data-driven phase, represented by the purple arrow in the figure, and explained in \Cref{sub:datadriven}. In this phase, we determine a linear operator that approximates the inverse radiative transfer process using training data. This step forms the first offline phase. Given an input spectrum $\bfy$, the operator, denoted as $\widehat{\bfZ}$, produces an estimate of the atmospheric scenario 
$\widehat{\bfx}$ using \Cref{eq:ddsol}. This phase is entirely data-driven and provides a preliminary solution to the atmospheric scenarios. The second step, shown by the blue arrow, incorporates a priori information $\bfx_a$ into the data-driven solution $\widehat{\bfx}$ through a Tikhonov regularization technique, as explained in \Cref{sub:aprioriapp}, resulting in $\bfx_{\bflambda}$. The inclusion of a priori information improves the initial estimate by applying additional constraints based on prior knowledge. The regularization parameters for this phase are estimated using a neural network, represented by the horizontal green arrow and discussed in \Cref{sub:lambdaopt}.

To train the neural network, we generated a training set of pre-computed optimal regularization parameters, as indicated by the vertical green arrow and detailed in \Cref{sub:lambdaopt}. This calculation forms the second offline phase. Together, these steps offer a comprehensive solution to the inverse problem, effectively combining data-driven methods with a priori information to produce estimates of atmospheric scenarios.

\section{Data organization}\label{sec:dataorg}
For the definition of our datasets we need the input variables describing the atmospheric scenarios and the output variables representing the corresponding FORUM simulated spectra.

The atmospheric state was build using the ECMWF (European Center for Medium range Weather Forecast) Re-Analysis ERA5 database  \cite{hersbach20}. Additional atmospheric gases concentrations of \ce{CO2}, \ce{CO}, \ce{CH4}, \ce{NO2}, \ce{HNO3}, \ce{SO2} were taken from the CAMS (Copernicus Atmosphere Monitoring Service) global Greenhouse Gas reanalysis (EGG4) \cite{egg2020}. The remaining atmospheric gases were obtained from the Initial Guess 2 (IG2) database \cite{remedios07}. The surface emissivity profiles were taken from the Huang database \cite{huang2016}, with an ad-hoc procedure to associate the most suitable profile to the time and geolocation \cite{sgheri2024}. Finally, the a priori data were generated by perturbing the true profiles according to a precision matrix $\bfS_a$, constructed based on error estimates provided by the UK Met Office for routine assimilation of IASI products into their operational Numerical Weather Prediction (NWP) system. 

For the spectral simulation, the radiative transfer problem was solved using the CLouds and Atmospheric Inversion Module (CLAIM), an advanced version of the inversion code used in \cite{sgheri2022}, based on the LBLRTM (Line By Line Radiative Transfer Model) forward model~\cite{clough05}. For the inversion, the CLAIM code uses the combination of the OE approach, the Gauss-Newton minimization method with the Levemberg-Marquardt technique and a final Tikhonov regularization step with the IVS (Iterative Variable Strength) method \cite{ridolfi2011,sgheri2020}. The FORUM instrument characteristics were taken from one of the two instrument concepts available in phase A/B1 of the FORUM E2E project.

Among the atmospheric variables, we focus on 5 key parameters for our reconstruction purposes: Earth surface temperature, air temperature, water vapor, ozone, and surface spectral emissivity. Earth surface temperature is a scalar, surface spectral emissivity is a vector depending on the wavenumber, normally approximated with a linear spline, while the other parameters are vertical profiles that can be defined in terms of atmospheric pressure levels, using a fixed grid with 41 levels.
However, the surface altitude of the geolocation determines the lowest pressure level for the radiative transfer, so that the actual number of pressure levels can be smaller. To standardize the data for our study, we require a uniform data dimension. We map the pressure range of each point onto a set of 41 pressure levels and then interpolate all other vertical profiles onto this grid. This ensures that while the actual pressure levels for each point may vary, the total number remains constant.  Consequently, the atmospheric state $\bfx$ is composed of $n=425$ components: one value for surface temperature, 41 components for the profiles of air temperature, water vapor, and ozone, and 301 values for surface spectral emissivity, defined over the spectral interval from 100 to 1,600 \si{cm^{-1}} and interpolated on a 5 \si{cm^{-1}} regular grid. 
Furthermore, the output variables are represented by the simulated FORUM spectra, which are given on $q = 4,\!049$ spectral points within the same spectral interval of 100 to 1,600 \si{cm^{-1}}. These spectra are generated by running LBLRTM on the corresponding input atmospheric scenarios.

For the purpose of our analysis, we collect information on atmospheric scenarios in January and July 2021 at 12:00 all over the world, focusing only on clear sky conditions, employing the same selection process established in \cite{sgattoni2024}.
To provide further details, our dataset encompasses information spanning the entire globe, utilizing a grid system with longitude and latitude increments of $10^\circ$ and $5^\circ$ respectively. To manage the dataset's size, we restrict the data to the initial $20$ days of both July and January and for each geographical location, we identify the first and second available clear sky days in the database. Consequently, not all geographical points in our dataset feature a clear sky scenario.

\begin{figure}[H]
\begin{center}
\includegraphics[width=1.
\textwidth]{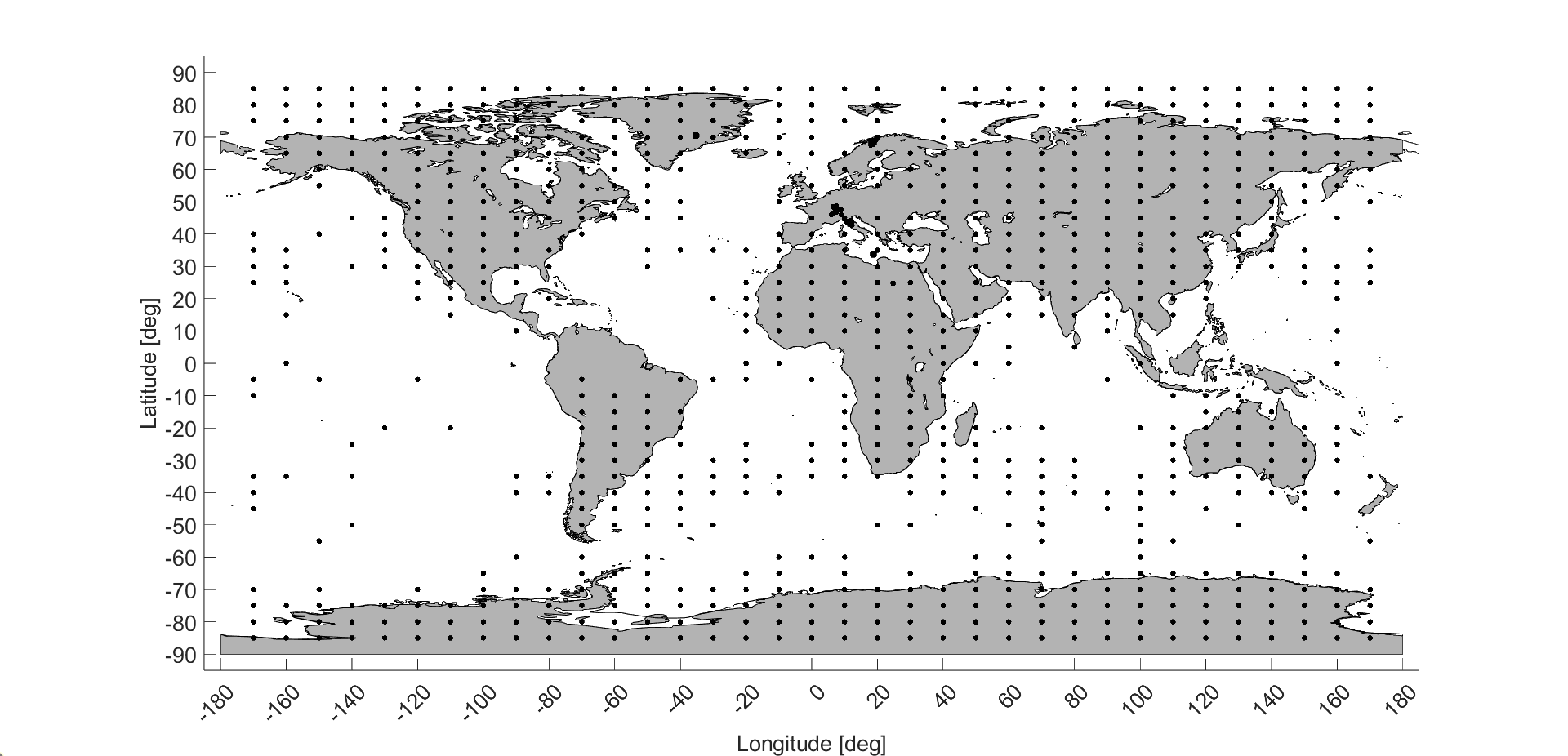}
\caption{Geolocations of cases in TN1, represented by black dots for both January and July.}\label{fig:tn1}
\end{center}
\end{figure}

We organize these data into different sets. First, we create training set 1 (TN1), containing all the data related to the first clear sky day of both January and July. We enrich this set with additional points corresponding to well-known and typical locations with different climates and soil types, i.e., the Sahara desert, the city of Florence, the Black Forest, Greenland, Finland, Sweden, and the Mediterranean Sea. 
TN1 is represented by the input set $\bfX \in \mathbb{R}^{n\times m}$ of atmospheric scenarios and the output set $\bfY \in \mathbb{R}^{q\times m}$ of spectra, with $m=1,\!708$ observations illustrated in Figure \ref{fig:tn1}.

Secondly, we use the remaining points corresponding to the second clear sky days in both January and July to construct two separate test sets, TS1 and TS2. We divide these points based on their indices, assigning odd indices to TS1 and even indices to TS2.

\begin{figure}[H]
\begin{center}
\includegraphics[width=1.
\textwidth]{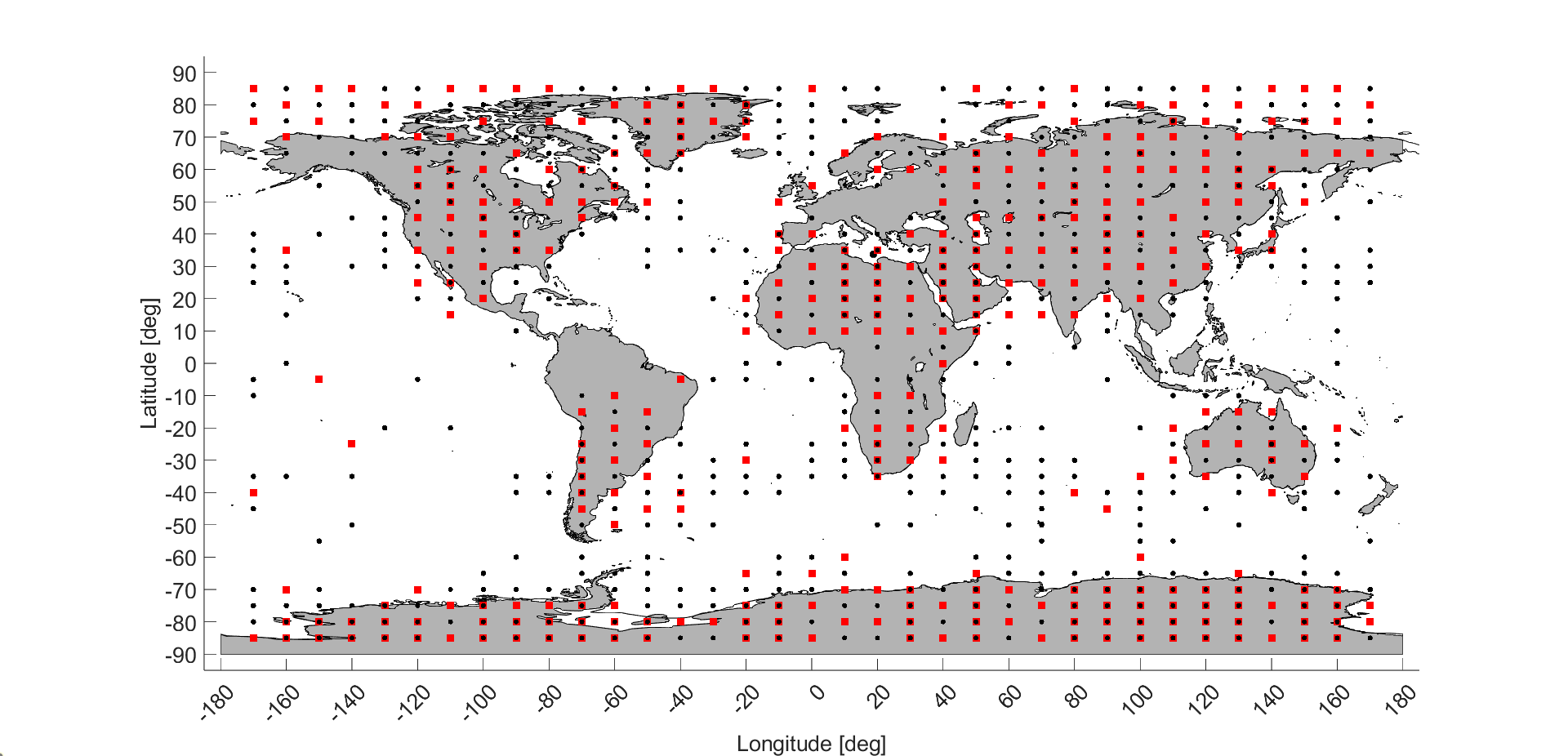}
\caption{Geolocations of cases in TS1 and TS2 for both January and July, with red squares for TS1 and black dots for TS2.}\label{fig:ts}
\end{center}
\end{figure}

As a result, TS1 and TS2 consist of the input sets $\bfX_{\text{test}1} \in \mathbb{R}^{n \times m_{t_1}}$ and $\bfX_{\text{test}2} \in \mathbb{R}^{n \times m_{t_2}}$, respectively, and the output sets $\bfY_{\text{test}1} \in \mathbb{R}^{q \times m_{t_1}}$ and $\bfY_{\text{test}2} \in \mathbb{R}^{q \times m_{t_2}}$, where, $m_{t_1} = 396$ and $m_{t_2} = 402$ represent the number of observations for each test set, respectively.
The composition of TS1 and TS2 is depicted Figure \ref{fig:ts}

Finally, we generate a priori information, as described at the beginning of this section, and associate it with the input variables in TS1 and TS2. This information is represented by $(\bfX_a)_{\text{test}1}$ and $(\bfX_a)_{\text{test}2}$, respectively, and is linked to the corresponding precision matrices $(\bfS_a)_{\text{test}1}$ and $(\bfS_a)_{\text{test}2}$.

In our inversion scheme, detailed in \Cref{sec:solsch}, we use TN1 in the first offline, fully data-driven training phase described in \Cref{sub:datadriven}. Then, TS1 is initially used to test this phase and later as a training set for the second offline training phase, as described in Sections \ref{sub:aprioriapp} and \ref{sub:lambdaopt}. Finally, TS2 is used to test the entire process chain.
Starting from the preliminary data-driven results in TS1 we compute the experimental precision matrix $\bfS_\bfx$ used in the subsequent phases.

In general, all variables involved in our simulations change significantly in magnitude, therefore, in order to have a similar scale, for the purpose of our analysis, we apply a normalization,  detailed in Appendix~\ref{appendix:norm}.
It is worth highlighting that the normalization map is defined in the training phase using only the training data and then used for the test data in the testing phases.

\section{Results}\label{sec:results}

In this section, we present the results from the various stages of our solution scheme. The experiments evaluate the proposed approach across each phase, specifically addressing the data-driven phase,  \Cref{sub:datadrivenres}, and the integration of a priori information with the estimation of regularization parameters, \Cref{sub:aprioriappres}. A detailed analysis of the experimental findings follows, highlighting the efficiency of our method across different scenarios. These results demonstrate the contribution of each component to the overall performance in atmospheric scenario reconstruction. Finally, in \Cref{sub:fullph} we compare our solution with a full-physics procedure.

\subsection{Fully data-driven results}\label{sub:datadrivenres}
\begin{wrapfigure}{r}{0.55\textwidth}
\begin{center}
\includegraphics[width=0.55
\textwidth]{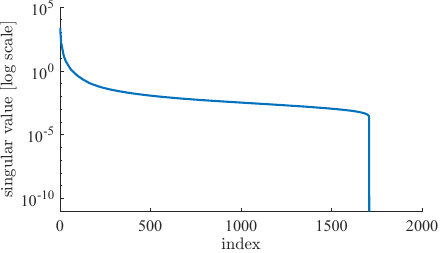}
\end{center}
\caption{Singular values representation of matrix $\bfY$. The singular values in decreasing magnitude. The rapid decline in the singular values suggests that the first few components capture most of the information, allowing for dimensionality reduction without significant loss of information.}\label{fig:fullMi_svd}
\end{wrapfigure}

In this section, we present the results obtained with the fully data-driven approach. As described in \Cref{sub:datadriven}, we train the operator $\widehat{\bfZ}$ using data from the TN1 database, as detailed in \Cref{sec:dataorg}. In equation~\eqref{eq:fullmi}, we apply the pseudoinverse operator with a tolerance of $10^{-6}$, treating all singular values below this threshold as zero. \Cref{fig:fullMi_svd} shows the singular values of matrix $\bfY$, which contains the $m$ spectra from TN1. The graph suggests that the chosen tolerance is appropriate and relatively conservative.

We use the resulting operator $\widehat{\bfZ}$ to test cases in TS1 database, as shown in \Cref{eq:ddsol}.
To evaluate the quality of the data-driven results, we apply the following quantifiers.

Given a single test spectrum $\bfy_j \in \bfY_{\text{test}1}$, in order to evaluate the performance of the inverse problem in reconstructing the corresponding parameter vector $\bfx_j$, we compute the error vector $\mathbf{e}_j$, whose components are given by 
$$e^i_{j} = x_j^i-(x_p)_j^{i}, \quad i=1,\dots,n,$$
where the prediction $(x_p)_j^{i}=\widehat x_j^i$ as in~\eqref{eq:ddsol}.
In order to evaluate any possible systemic bias in the error behavior, an analysis of the whole $m_{t_1}$ aggregated cases is carried out, computing the mean signed error $\mathbf{E}$ and the mean unsigned error $\mathbf{uE}$, whose components are:
$$E^i = \frac{1}{m_{t_1}} \sum_{j=1}^{m_{t_1}} e_j^{i}, \quad i=1,\dots,n,$$
$$(uE)^i = \frac{1}{m_{t_1}} \sum_{j=1}^{m_{t_1}} |e_j^{i}|, \quad i=1,\dots,n.$$
The first one is useful for evaluating a possible systematic bias in the error behavior, for example, whether a quantity is consistently over- or underestimated. The second one is more useful as a measure of overall accuracy.

\begin{figure}[H]
\begin{center}
\begin{tabular}{cc}
    \includegraphics[width=0.4\textwidth]{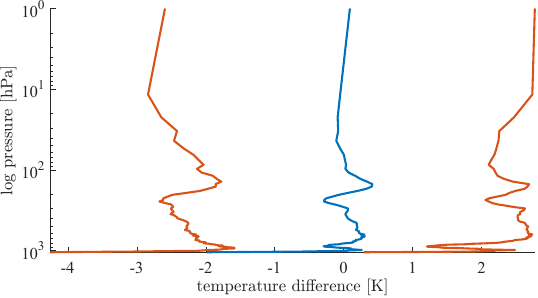} & 
    \includegraphics[width=0.4\textwidth]{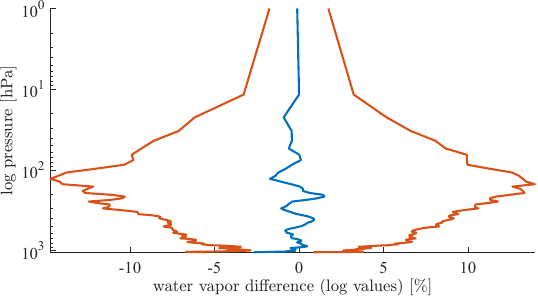} \\
    \includegraphics[width=0.4\textwidth]{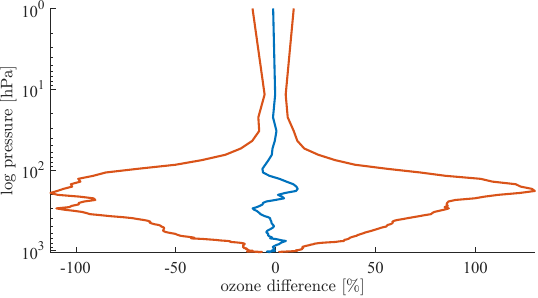} & 
    \includegraphics[width=0.4\textwidth]{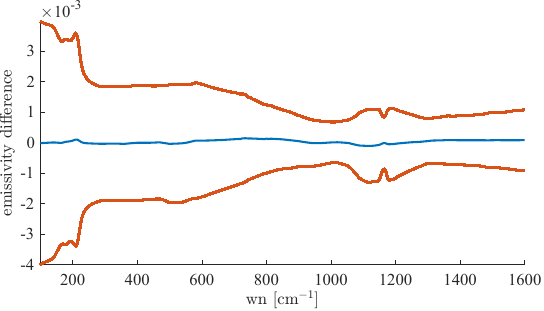} 
\end{tabular}
\end{center}
\caption{Results from the purely data-driven approach. Mean signed and unsigned reconstruction errors for the entire TS1 are shown in orange and blue, respectively. The panels, in order, correspond to temperature, water vapor, ozone, and surface spectral emissivity. The surface temperature difference is $5.901 \cdot 10^{-4} \pm 0.063$ \si{K}.}\label{fig:aggregated_ddsol}
\end{figure}

Figure~\ref{fig:aggregated_ddsol} presents the aggregated results for the entire TS1, which includes 396 cases. The figure displays both the mean signed errors, $\mathbf{E}$, and unsigned errors, $\mathbf{uE}$, in blue and orange, respectively, offering a comprehensive overview of the model's performance. The arrangement of the panels, starting from the top left and moving clockwise, reflects the reconstruction errors for the 5 key quantities of interest: surface temperature, atmospheric temperature, water vapor, ozone, and surface spectral emissivity. Notably, logarithmic values are employed for water vapor due to the substantial variations across pressure levels, a strategy that improves the training process. Overall, the signed errors tend to approach zero, remaining well within the unsigned error bars, which suggests that our method is generally unbiased across the different variables.

\begin{figure}[H]
\begin{center}
\begin{tabular}{cccc}
\includegraphics[width=0.4\textwidth,]{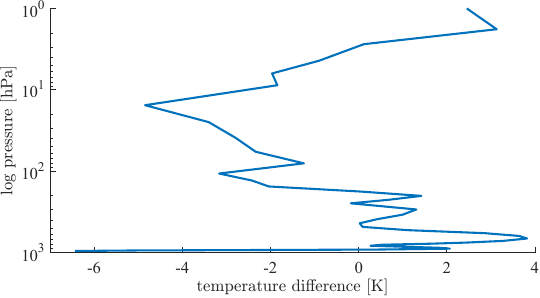} & 
\includegraphics[width=0.4\textwidth,]{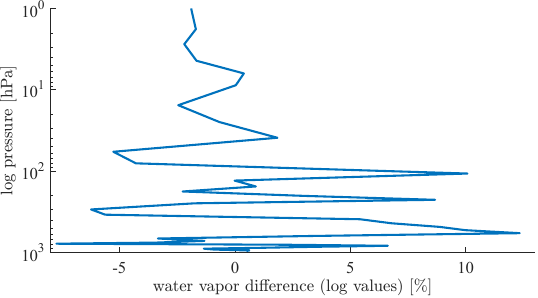} \\
\includegraphics[width=0.4\textwidth,]{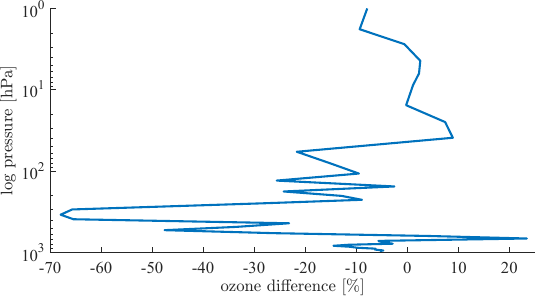} & 
\includegraphics[width=0.4\textwidth,]{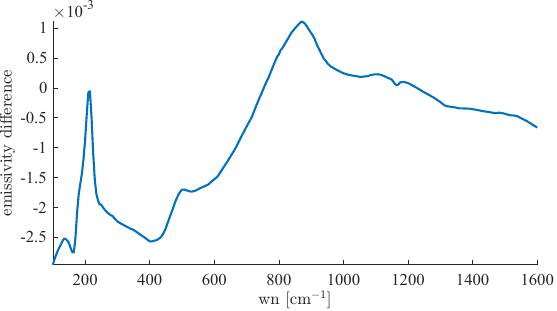} 
\end{tabular}
\end{center}
\caption{Results from the purely data-driven approach. Reconstruction unsigned error for case 25 in TS1. The panels, in order, correspond to temperature, water vapor, ozone, and surface spectral emissivity. The  surface temperature difference is $-2.244\cdot 10^{-2}$ \si{K}.}\label{fig:case25_ddsol}
\end{figure}

In Figure \ref{fig:case25_ddsol}, we present the pointwise results for a specific case in TS1, selected using a random number generator. The figure, composed of five panels similar to the previous one, shows reasonable unsigned error profiles for the quantities of interest. However, some unphysical oscillations and unusual local behavior are observed, particularly in the lower atmospheric layers for ozone and water vapor, and to a lesser extent for temperature. While these oscillations remain within an acceptable range, their frequency and amplitude suggest non-physical behavior.
In summary, based on the previous figures, the fully data-driven method generally provides high-quality reconstructions, assuming the training set is representative of the test set. However, the method fails in some individual cases. In other instances, strange local behavior and irregularities appear within otherwise accurate reconstructions.
This fully data-driven approach is already capable of providing useful results in most cases, though it could be improved. Incorporating a priori information, as suggested in \Cref{sub:aprioriapp}, may be beneficial.

\subsection{Data-driven results with a priori correction} \label{sub:aprioriappres}
A priori information consists of atmospheric data known from climatology or external sources related to a specific geolocation and time. These data have been specifically selected for each test case and contain information for surface temperature, temperature, water vapor, ozone, and surface spectral emissivity, as described in Section~\ref{sec:dataorg}. This information is collected in the vector $\mathbf{x}_a$ in Equation~\eqref{eq:apsol} representing an estimate of our solution. The atmospheric variability of these variables is represented in the related precision matrix $\mathbf{S}_a$ constructed from the standard deviation and the correlation length of those variables provided by the UK Met Office for routine assimilation of IASI products into their operational Numerical Weather Prediction (NWP) system. 

At this point, the challenge is to determine the appropriate influence of the a priori information on our solution, which corresponds to selecting good values for $\bflambda$ in the minimization problem given by~\eqref{eq:ap}. It is important to note that the influence of the a priori information may vary for each of the five variables of interest, as well as for each component within these variables. In our experiments, we opt for a constant regularization parameter vector for each variable, resulting in a final minimization problem that involves determining five constant concatenated vectors. To address this, we analyze all $m_{t_1}$ cases in TS1 and, for each case, we use \Cref{eq:apsol} to solve \Cref{eq:tf} as detailed in Sections~\ref{sub:aprioriapp} and~\ref{sub:lambdaopt}. For this purpose, we employ an optimization algorithm based on the interior-point method, using a tolerance of $10^{-5}$ for both the objective function in Equation~\eqref{eq:tf} and the solution, i.e., the optimal regularization parameters. We initialize the algorithm with random vectors, each component ranging between 0 and 1.
Specifically, we use Matlab's constrained optimization method \texttt{fmincon} with default settings, which is commonly used to find the minimum of constrained nonlinear multivariable functions.

It is important to note that the optimization process using the interior-point algorithm is repeated multiple times (10) with different initial conditions. This approach is known as the multi-start method and helps mitigate the risk of convergence to local minima and increases the likelihood of finding the global optimum. By running the algorithm multiple times, we can ensure the robustness of the solution and verify that the results are not overly dependent on the choice of initial values. For a detailed discussion of the bilevel optimization strategy for the five regularization variables and the aggregation of results via singular value decomposition (SVD), please refer to the \Cref{appendix:svd}.

\begin{figure}[H]
\begin{center}
\includegraphics[width=1.
\textwidth]{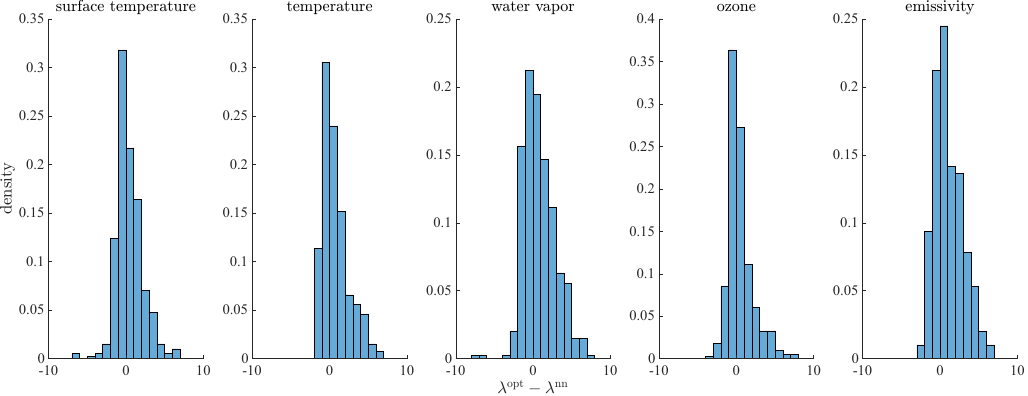}
\end{center}
\caption{Relative difference between the optimal regularization parameters $\bflambda^{\text{opt}}$ and the estimated ones $\bflambda^{\text{nn}}$.  The panels correspond, in order, to the surface temperature component, temperature component, water vapor component, ozone component, and surface spectral emissivity component.}\label{fig:nnperf}
\end{figure}

At this point, for cases in TS1, given the data-driven solutions $\widehat{\bfx}_j$, the corresponding a priori data $(\bfx_a)_j$, and the optimal regularization parameters $\lambda^{\text{opt}}_j$, for $j=1,\dots,m_{t_1},$ we can train a neural network as detailed in Equations~\eqref{eq:ffnn1} and~\eqref{eq:ffnn} in \Cref{sub:lambdaopt}.
We employ a feedforward neural network for all five variables, consisting of three hidden layers with sizes of 15, 10, and 5 units, respectively. In the hidden layers, we use a ReLu activation function. The training is performed using the Levenberg-Marquardt optimization technique.
Figure~\ref{fig:nnperf} presents histograms illustrating the network's performance in estimating the regularization parameters for cases in TS1. The figure is composed of five panels, each corresponding to one of the five variables. It shows the relative difference between the previously computed optimal regularization parameters $\bflambda^{\text{opt}}$ and the ones estimated by the network, denoted by $\bflambda^{\text{nn}}$.

\begin{figure}[H]
\begin{center}
\includegraphics[width=0.8
\textwidth]{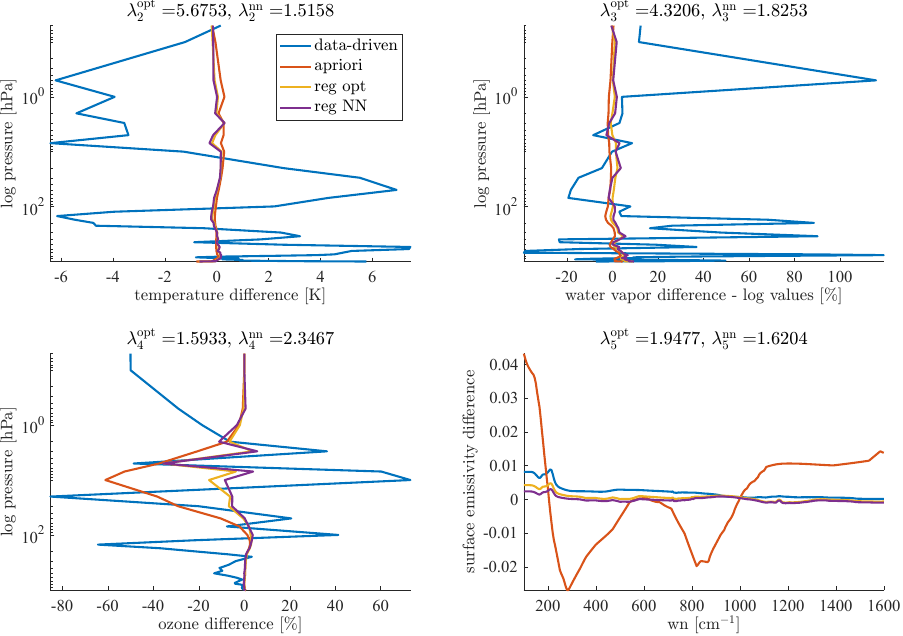}
\end{center}
\caption{Comparison for case 256 between the unsigned reconstruction errors relative to the data-driven solution (blue), the a priori solution (red), the regularized solution using the optimal parameters (yellow), and the regularized solution using the estimated parameters (violet). The panels, in order, correspond to temperature, water vapor, ozone, and surface spectral emissivity. The respective regularization parameter values are reported at the top of each panel. For surface temperature, $\lambda_1^{\text{opt}} =5.6753$ and $\lambda_1^{\text{nn}} = 2.8512$.}\label{fig:nn256}

\end{figure}

To closely examine the neural network's performance, we focus on a single representative case from TS1. Figure~\ref{fig:nn256} refers to case 256 and contains four panels corresponding to temperature, water vapor, ozone, and surface spectral emissivity. These panels show the unsigned errors between the true solution and the following: the fully data-driven solution (in blue), the a priori solution (in red), the regularized solution using the optimal regularization parameters (in yellow), and the regularized solution using the estimated regularization parameters (in violet). The exact values of the regularization parameters are reported at the top of each panel. For the sake of clarity, we refer to them, using the notation $\lambda^i,$ where the superscript $i=1,\dots,5,$ does not refer to the component index but to the variable index.
As shown, the quality of the solution does not degrade when using the neural network's estimated regularization parameters.

\

\begin{figure}[H]
\begin{center}
\begin{tabular}{cc}
    \includegraphics[width=0.4\textwidth]{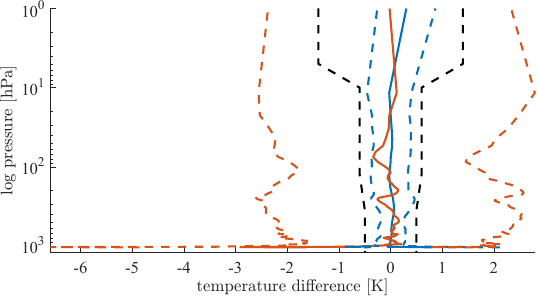} &
    \includegraphics[width=0.4\textwidth]{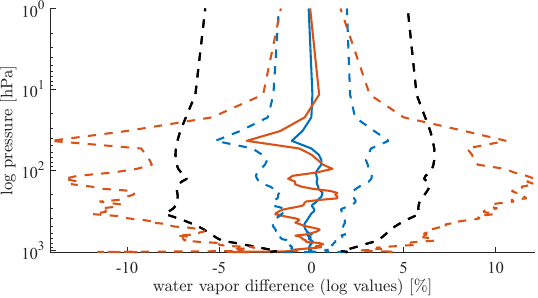}\\
    \includegraphics[width=0.4\textwidth]{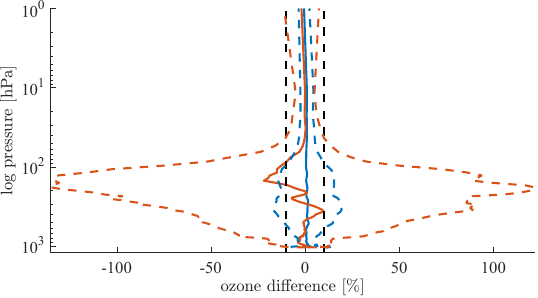} &
    \includegraphics[width=0.4\textwidth]{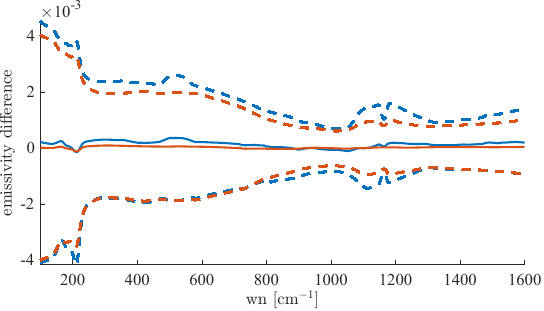}
\end{tabular}
\end{center}
\caption{Mean signed and unsigned reconstruction errors for the entire TS2 dataset are shown using bold colored lines for signed errors and dashed colored lines for unsigned errors. Errors related to the data-driven solution are represented in orange, while those for the regularized solution are in blue. Black dashed lines indicate the a priori errors. Each panel corresponds, in order, to atmospheric temperature, water vapor, ozone, and surface spectral emissivity. The a priori error $\pm 0.05$ for surface emissivity is omitted for clarity. The surface temperature difference for the data-driven solution is $-0.002 \pm 0.057$ \si{K} and for the regularized solution is $-0.01 \pm 0.0714$ \si{K}.}\label{fig:final_aggregated}
\end{figure}

\begin{figure}[H]
\begin{center}
\begin{tabular}{cc}
    \includegraphics[width=0.4\textwidth]{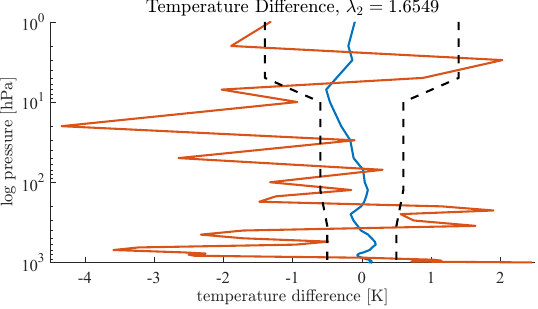} &
    \includegraphics[width=0.4\textwidth]{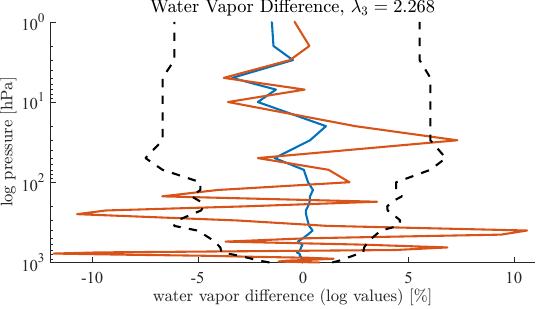}\\
    \includegraphics[width=0.4\textwidth]{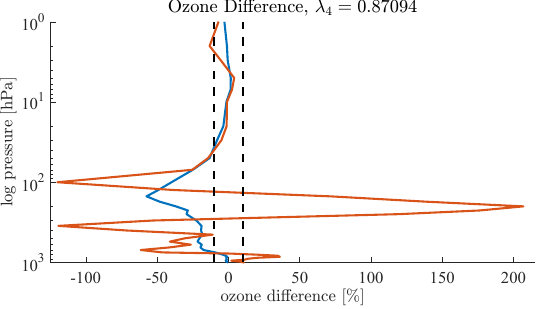} &
    \includegraphics[width=0.4\textwidth]{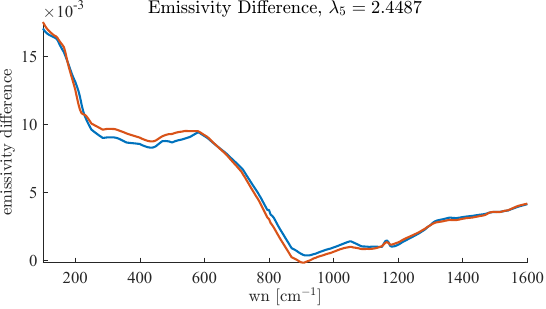}
\end{tabular}
\end{center}
\caption{Signed errors with respect to the true values for case 15 in TS2, with data-driven solutions shown in orange and regularized solutions in blue.
The dashed lines represent the a priori errors. The panels, in order, correspond to atmospheric temperature, water vapor, ozone, and surface spectral emissivity. The a priori error $\pm 0.05$ for surface emissivity is omitted for clarity. The surface temperature difference for the data-driven solution is $0.019$ \si{K} and for the regularized solution is $-0.021$ \si{K}. The values of the estimated regularization parameter components are reported at the top of each panel, with $\lambda_1= 1.0993$.}\label{fig:final_case15}

\end{figure}

At this point, we have all the elements to test the entire solution scheme on a set of new cases. In particular, we consider the 402 points in TS2. Figures~\ref{fig:final_aggregated} and~\ref{fig:final_case15} show a comparison between the fully data-driven phase and the complete solution procedure. Specifically, Figure~\ref{fig:final_aggregated} displays the mean signed and unsigned errors in bold and regular font, respectively, for the fully data-driven solution in orange and the regularized solution in blue. The five panels correspond to the usual five variables in the standard order. Once again, we observe that the full method is unbiased and that regularization produces a significant improvement in the solution.

Next, we examine individual cases to assess whether the method now performs better even at a pointwise level. Figure~\ref{fig:final_case15} shows the unsigned errors for the data-driven solution in blue and the regularized solution in orange for a single random case in TS2. The values of the estimated regularization parameters are reported at the top of each panel using the same notation as in the previous figure. As you can see, the regularization greatly improves the results, leading to a method that is quite accurate even for individual retrievals.

\subsection{Comparison with a full-physics method}\label{sub:fullph}
Next, we compare our method with a full-physics approach, specifically the CLouds and Atmospheric Inversion Module (CLAIM), as described in Section~\ref{sec:dataorg}. In terms of computational time, our method outperforms the traditional approach. The CLAIM method requires approximately between 1 and 2 hours to perform a single retrieval, depending on the number of Gauss-Newton iterations needed for convergence. The bottlenecks are the cross-section evaluation via the spectroscopic database and the convolution with the ISRF that must be repeated for the Jacobian of each retrieval parameter, or about 450 times for each Gauss-Newton step. In contrast, our method takes 0.178
seconds for the initial training phase and 128.984
seconds to train the neural network during the offline stage. The online stage requires about 0.0008 seconds for the fully data-driven solution of a single case, approximately 0.007 seconds to estimate the regularization parameters for a single case, and around 0.05 seconds for the entire solution scheme for a single case.

Of course, such a fast method cannot compete in terms of precision and accuracy with the full-physics method. To evaluate the performances of our method on the full TS2, we use the chi-square statistics, which measure the goodness of the spectrum reconstructed with our solution versus the full-physics solution \cite{rodgers2000}.

\begin{figure}[H]
\begin{center}
\includegraphics[width=0.6
\textwidth]{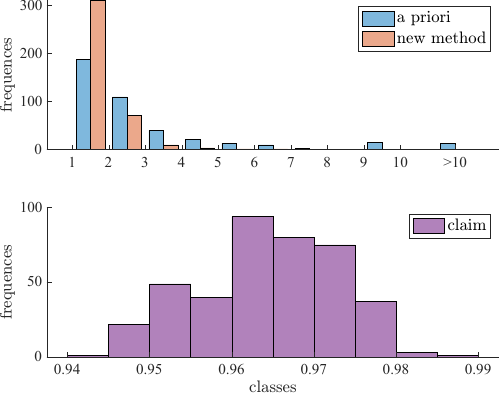}
\end{center}
\caption{Binned $\chi^2_r$ distribution. Top panel: A priori (blue) and results of new method (orange). Bottom panel: full-physics method.}\label{fig:chi2_comparison}
\end{figure}

More precisely, we use the reduced $\chi^2_r$ statistics calculated as follows:
\begin{equation}
    \chi^2_r = \frac{\big(\bfy-\bfF(\bfx)\big)\t\bfS_\bfy^{-1}\big(\bfy-\bfF(\bfx)\big)}{q},
\end{equation}
where, as introduced in the previous sections, $\bfy$ is the measured spectrum, $\bfF(\bfx)$ the simulated spectrum, $\bfS_\bfy$ the measurements precision matrix and $q$ the dimension of the spectrum.
The expected value of the $\chi^2_r$ for a spectrum that is compatible with the instrumental noise is $1$, and $(q-m)/q\sim 0.9$ for the reconstruction after the minimization. Of course, the full-physics method performs better, as it was expected. However, we also calculated the $\chi^2_r$ statistics obtained when using the a priori as a solution. The results are presented in Figure~\ref{fig:chi2_comparison}. From the figure, we see that our solution improves the given a priori, though the a priori were calculated as a background error, hence with a reasonably small $\bfS_a$. It is important to get accurate a priori also as a starting point for the full-physics solution because the bias of a solution depends on the distance between the a priori and the truth \cite{merchant2020,sgheri2024}. If the a priori were taken from the climatology, the average error would be much larger. 

\section{Conclusions}\label{sec:concl}

In this work, we present a novel data-driven approach for approximating the inverse mapping ``$\bfF^{-1}$'' in the retrieval problem, aiming for a solution that is both computationally efficient and robust. The use of simulated FORUM spectrometer measurements as training data, combined with the integration of climatological information, allows us to address the ill-posed nature of the inverse problem. A key feature of our method is the initial purely data-driven phase, where we approximate ``$\bfF^{-1}$'' without relying on any prior information, ensuring flexibility and generalization across different scenarios. However, since the inverse problem is inherently ill-posed, approaches that do not apply regularization techniques are unable to accurately recover the profiles. The second key innovation lies in the use of a neural network to estimate the optimal regularization parameters during the retrieval process to incorporate prior information.
Although our approach does not match the precision of full-physics retrieval methods, it offers near-instantaneous results, making it a valuable tool for applications where speed is crucial. Additionally, the proposed method's ability to provide a priori values that are statistically closer to the true atmospheric state suggests that it can reduce the bias in the inverse problem solution typically introduced by errors in the a priori data. Overall, our method provides a practical alternative to traditional retrieval approaches, especially in scenarios where computational efficiency and the ability to handle large datasets are essential.

\section{Acknowledgments}
INdAM-GNCS supported the first author under Bando di concorso a n.30 mensilità di Borse di studio per l’estero A.A. 2022-2023. This work was partially supported by the National Science Foundation (NSF) under grant [DMS-2152661] for M. Chung. Any opinions, findings, conclusions, or recommendations expressed in this material are those of the authors and do not necessarily reflect the views of the National Science Foundation. 

\printbibliography

\begin{appendix}

\section{Data Normalization}\label{appendix:norm}
Consider our matrix of input data $\bfX\in \mathbb{R}^{n\times m}$ and output data $\bfY \in \mathbb{R}^{q \times m}$. The columns of $\bfX$ and $\bfY$ correspond to different observations, and the rows to the different variables in each observation.
\par The variables in the training sets $\bfX$ and $\bfY$ may differ significantly in magnitude. Consequently, it is often helpful to normalize the data so that all quantities of interest have a similar scale, allowing us to better exploit the correlations between each variable.
\par We employ the following \textit{center-and-scale} normalization procedure, as follows:
\begin{equation}
\widehat\bfX = \text{diag}(\sigma(\bfX))^{-1}\left(\bfX - \overline{\bfx}\cdot \mathbf{1}\t\right),\qquad \widehat\bfY = \text{diag}(\sigma(\bfY))^{-1}\left(\bfY - \overline{\bfy}\cdot \mathbf{1}\t\right), 
\end{equation}
where $\overline{\bfx},\,\overline{\bfy}$ and $\sigma(\bfX),\,\sigma(\bfY)$ denote the vectors containing the mean and standard deviation of each row of $\bfX,\,\bfY$ respectively, and $\mathbf{1}$ is the appropriately-sized vector containing all ones.
\par One may un-normalize the observations in a natural way, with the inverse maps given by:
\begin{equation}
\bfX = \text{diag}(\sigma(\bfX))\widehat\bfX + \overline{\bfx}\cdot \mathbf{1}\t,\qquad \bfY = \text{diag}(\sigma(\bfY))\widehat\bfY + \overline{\bfy}\cdot \mathbf{1}\t. 
\end{equation}
In practice, we define these maps during the training phase, using the training data only. The maps are then saved and re-used, with no changes, on the test data during the testing phase. For this normalization procedure to be effective, it is, therefore, necessary that the space spanned by the training sets is sufficiently rich to represent the data contained in the test sets.

\section{Methodology for Independent Minimization and Aggregation of Regularization Variables} \label{appendix:svd}

A key aspect of our optimization approach in \Cref{sub:aprioriappres} is addressing the varying magnitudes of the five regularization variables. We perform the outer minimization of~\eqref{eq:tf} independently for each variable, while still allowing the others to vary in each case, as they are not fully independent.
More precisely, we change the target function each time for each variable, i.e., the outer problem in~\eqref{eq:tf}, while keeping the inner problem in~\eqref{eq:apsol} fixed at its original dimension. 
At the end of each optimization process, we obtain five distinct vectors for each case. Each vector contains five components, representing the optimized values for all five variables, but with the minimization focused specifically on one variable at a time.
To aggregate the results and obtain a single vector that performs well across all variables we collect the five vectors into a square matrix $\bfW$. We then perform the singular value decomposition (SVD) on this matrix and derive the optimal solution using a linear combination of the first two singular values
$\sigma_1$ and $\sigma_2$ and the corresponding left singular vectors, where the coefficients are chosen empirically:

\begin{equation}\label{eq:lopt_svd}
    \textbf{W} = \textbf{USV}\t, \quad 
    \bflambda^{\rm opt} = \big|\tfrac{1}{4} \sigma_1\bfu_1 +\sigma_2\bfu_2\big|,
\end{equation}

\begin{figure}[h]
\begin{center}
\includegraphics[width=0.55
\textwidth]{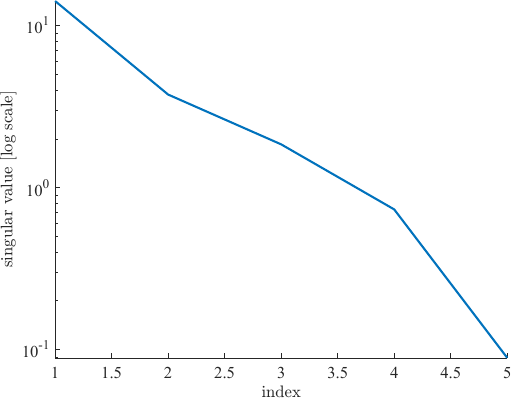}
\caption{Typical $\textbf{W}$ singular values behavior. The singular values, arranged from largest to smallest, reflect how much each component contributes to explaining the variance within the original matrix. The steep drop-off in singular values indicates that the first two components contain the bulk of the information, making it possible to reduce dimensionality with minimal data loss. Smaller singular values likely represent noise and can be omitted to achieve data compression.}\label{fig:svd255}
\end{center}
\end{figure}

To justify this choice, \Cref{fig:svd255} shows the typical decrease of the singular values of matrix $\bfW$. Similar patterns are observed in all cases, where the first two singular values capture the most informative components of the data.
\par In order to better understand the formulation, we analyze the relationship between matrix $\bfW$ and $\bfW\bfW\t$.

By definition, all elements of $\bfW$ are non-negative, so this also applies to $\bfW\bfW\t$. Since $\bfW\bfW\t$ is non-negative, by the weak Perron-Frobenius theorem (see, for instance, Chapter 8 of \cite{meyer2000}), the eigenvector $\bfu_1$ corresponding to the largest eigenvalue $\mu_1$ of $\bfW\bfW\t$ can be chosen so that all its components are not negative. Moreover, since we verified that $\mu_1$ is simple and that positive left and right eigenvectors exist for $\mu_1$, by the Gantmacher theorem \cite{gantmacher1959}, $\bfW\bfW\t$ is also irreducible, so we can apply the strong Perron-Frobenius theorem that guarantees that $\bfu_1$ has strictly positive components. By definition, $\bfu_1$ is also the left singular vector of $\bfW$ corresponding to the singular value $\sigma_1=\sqrt{\mu_1}$.
\par Using again the Perron-Frobenius theorem, we know that $\bfu_1$ is the only eigenvector with all positive components, meaning that $\bfu_2$ is not entry-wise nonnegative. Given that the $\bfu_i$ vectors have unit norm and that $\sigma_1$ is larger than $\sigma_2$ (Fig.~\Cref{fig:svd255}), the linear combination of the first two singular vectors, each weighted by its respective singular value, will statistically tend to have positive components. However, we observed empirically that better results are achieved by adding a coefficient of $1/4$ to $\bfu_1$. Therefore, as positive values are required for $\bflambda^{\rm opt}$, we included the absolute value in~\Cref{eq:lopt_svd}.

\end{appendix}

\end{document}